\documentclass[prl, twocolumn, aps,amsmath,amssymb,amsfonts, superscriptaddress]{revtex4-2}
\usepackage{graphicx} 
\usepackage{graphicx}
\usepackage{graphics}
\usepackage{amsmath}
\usepackage{amsthm} %
\usepackage{amssymb}%
\usepackage{amsfonts}
\usepackage{amssymb}
\usepackage{epstopdf}
\usepackage{makeidx}
\usepackage{epsfig}
\usepackage{amsfonts}
\usepackage{bm}
\usepackage{color}
\usepackage{xcolor}
\usepackage{bbold}
\usepackage{afterpage}
\usepackage{empheq}
\usepackage{hyperref}
\usepackage{booktabs}
\usepackage{subfigure}
\usepackage{makecell}

\begin{document}
\title{Zeeman-activated Berry curvature  magnetotransport \\ from the bulk of  non-magnetic metals with inversion symmetry}

\author{Rhonald Burgos Atencia}
\affiliation{Dipartimento di Fisica "E. R. Caianiello", Universit\`a di Salerno, IT-84084 Fisciano (SA), Italy}

\author{Antonio Vecchione}
\affiliation{CNR-SPIN, c/o Universit\`a di Salerno, IT-84084 Fisciano (SA), Italy}

\author{Denys Makarov}
\affiliation{Helmholtz-Zentrum Dresden-Rossendorf e.V., Institute of Ion Beam Physics and Materials Research, 01328 Dresden, Germany}

\author{Carmine Ortix}
\affiliation{Dipartimento di Fisica "E. R. Caianiello", Universit\`a di Salerno, IT-84084 Fisciano (SA), Italy}
\affiliation{CNR-SPIN, c/o Universit\`a di Salerno, IT-84084 Fisciano (SA), Italy}

\begin{abstract}
The Berry curvature (BC) -- a quantity encoding the geometry of electronic wavefunctions -- governs various electronic transport effects in quantum materials. In magnetic systems, the BC is reponsible for the intrinsic part of the anomalous Hall conductivity. Local concentrations of BC in non-centrosymmetric materials can lead instead to the quantum nonlinear Hall effect. 
Here, we argue that the bulk of non-magnetic metals with inversion symmetry – systems where the BC is forced to vanish at any momentum -- can be endowed with substantial concentrations of BC even with an infinitesimally small Zeeman coupling. This Zeeman-activated BC, \emph{independent} of the magnetic field strength and instead related to the degree of non-parabolicity of the electronic bands, couples to the electronic orbital motion to generate a negative longitudinal magnetoresistance that scales with the relaxation time as the Drude resistivity. We show that the Zeeman-actived BC and the related  intrinsic negative magnetoresistivity are generic: they appear in \emph{all} centrosymmetric point groups and can occur both in topological and conventional conductors.
\end{abstract}

\maketitle

\paragraph{Introduction --}
The electronic Bloch waves of a solid 
define a mapping from the Brillouin zone to a complex space equipped with a geometric structure that is encoded in a quantum geometric tensor~\cite{pro80}. The imaginary part of this tensor corresponds to an emergent field in momentum space called Berry curvature (BC). In metallic magnets, the flux of the BC through the Fermi surface is finite and regulates the intrinsic part of the anomalous Hall conductivity~\cite{Haldane2004,Nagaosa2010}. Anomalous Hall effect measurements thus provide an electronic transport footprint of the electrons quantum geometry. 
In non-magnetic metals, the flux of the BC is forced to vanish by time-reversal symmetry. However, a segregation of positive and negative regions of BC is allowed in materials with non-centrosymmetric crystalline arrangements~\cite{XiaoDi2010}. 
These local concentrations of BC can result in a BC dipole that governs the so-called nonlinear Hall effect with time-reversal symmetry~\cite{Sodemann2015,Carmine2021,Du2021}. A closely related disorder-induced nonlinear Hall effect~\cite{Du2019}, which has been shown to appear at the surfaces of non-magnetic materials with trigonal symmetry, stems from an higher order moment of the Berry curvature  dubbed BC  triple~\cite{He2021,Makushko2024}.

In nonmagnetic metals with centrosymmetric crystalline arrangements  the electronic BC is forced to vanish at any momentum. 
Nevertheless, externally applied electromagnetic fields can be used to break either inversion or time-reversal symmetry and thus generate substantial BC concentrations. This typically occurs in 
two-dimensional materials with valley degrees of freedom where a protected level crossing is transmuted into an avoided level crossing by the external field,  
and BC hot-spots -- regions in close proximity to  near degeneracies where the Bloch waves are rapidly changing -- consequently appear. In bilayer graphene, for instance, an external electric field perpendicular to the layers generates a spectral gap reducing the point group symmetry from ${\mathcal D}_{3d}$ to ${\mathcal C}_{3v}$,  and endows the materials with a BC leading to high-frequency rectification~\cite{Isobe2019}, and a BC dipole if an additional uniaxial strain field is present~\cite{Battilomo2019}. 
In single-valley bulk three-dimensional solids a similar generation of substantial BC concentrations is not predicted, at least in the weak-field limit. 

In this Letter, we will demonstrate that contrary to these expectations, substantial BC concentrations arise in strongly spin-orbit coupled three-dimensional materials in the presence of an even infinitesimally small Zeeman coupling. This Zeeman-activated BC is \emph{independent} of the magnetic field strength and Land\'e $g$-factor, with its strength that is only related to the degree of non-parabolicity of the electronic bands. Additionally, it 
appears in \emph{all} centrosymmetric point groups both for topological and conventional electronic bands. 
We will further show that the Zeeman-activated BC couples via the anomalous velocity~\cite{Sundaram1999,GaoYang2014,GaoYang2015}  to the orbital motion of Bloch electrons, and generate a weak-field negative longitudinal magnetoresistance (NLMR): a decrease  in the electrical resistance when the driving electric field ${\bm E}$ is parallel to the externally applied magnetic field ${\bm B}$. Contrary to the NLMR observed in materials with strongly anisotropic dispersions~\cite{Pippard1964} or due to extrinsic mechanisms, including  anisotropic scattering \cite{ArgyresPN1956,SondheimerEH1962,JonesMC1967,StroudD1976} and inhomogeneous boundary effects \cite{MarkLee1994,MillerDL1996}, the NLMR we predict here violates the so-called Kohler relation~\cite{Kohler1938,PalHK2010}. We find in fact  that the ratio between the finite field resistivity $\rho({\bm B})$ and the Drude resistivity $\rho_0$ is independent of $\omega_c \tau $, with $\omega_c$ indicating the cyclotron frequency and $\tau$ the relaxation time. Put differently, the predicted negative magnetoresistivity $[\rho(B) - \rho(0)]/\rho(0)$  is an intrinsic quantity independent of disorder. 

The weak-field NLMR due to the Zeeman-activated BC we put forward here, provides an explanation for the NLMR experimentally observed in Bi$_2$Se$_3$ thin films~\cite{Wiedmann2016}, and generally represents  an  analog for centrosymmetric materials of the weak-field NLMR appearing in a number of non-centrosymmetric systems, 
such as Dirac and Weyl semimetals~\cite{Huang2015,Arnold2016,YanBinghai2017,Armitage2018,Varma2024}, which possess non-trivial quantum geometric properties already in the complete absence of external fields~\cite{SonDT2013,SpivakB2016,Sekine2017,NandyS2017,SharmaGargee2017,YangGao2017,JohanssonAnnika2019}.

\paragraph{Zeeman-activated Berry curvature --} We first discuss the generic appearance of a Zeeman-activated BC that is independent of the magnetic field strength. To make things concrete, let us consider a crystalline system with the maximally symmetric point group equipped with inversion symmetry: the cubic ${\mathcal O}_h$ group. Additionally, we will consider that in our single valley system  the low energy bands are centered around the $\Gamma$ point of the Brillouin zone. Due to the concomitant presence of time-reversal symmetry $\Theta$ and inversion symmetry ${\mathcal I}$, all the electronic states are guaranteed to be twofold degenerate, with each pair of bands that are Kramers' related. The effective long wavelength Hamiltonian for a single Kramers' pair can be written as ${\mathcal H}({\bf k})= d_0{(\bf k}) \sigma_0 + \vec{{\bf d}}({\bf k}) \cdot \vec{\boldsymbol{\sigma}}$, with the Pauli matrices $\vec{\boldsymbol \sigma}=(\sigma_x,\sigma_y,\sigma_z)$ acting in spin space and the crystalline momentum polynomials $\vec{{\bf d}}({\bf k})$ constrained by point group symmetries. Inversion symmetry, for instance, requires that only terms quadratic in momentum ${\bf k}$ are allowed. Together with the fact that the Pauli matrices transform in the $T_{1u}$ three-dimensional irreducible representation of the ${\mathcal O}_h$ group, we then obtain the effective Hamiltonian ${\mathcal H}_0({\bf k})= \hbar^2  \left(k_x^2+k_y^2+k_z^2 \right) \sigma_0 / (2 m^{\star})$ which implies that the two degenerate bands are perfectly parabolic and isotropic with effective mass $m^{\star}$. 

\begin{table}[t]
\centering
\resizebox{\columnwidth}{!} {
\begin{tabular}[t]{|c|c|c|c|c|}
\hline
$\bm d /\mathcal{M}$  & $\mathcal{M}^{\prime}_x=\mathcal{M}_x \Theta $  & $\mathcal{M}_y$ & ${\mathcal M}^{\prime}_z=\mathcal{M}_z \Theta $ & Allowed \\
\hline \hline
$ d_x $   & $k_xk_y, k_xk_z$        &  $k_xk_y, k_zk_y$     &  $k_xk_y, k^2_i $        & $ak_xk_y$          \\ \hline \hline
$ d_y  $  & $b, k^2_i , k_yk_z$      &  $b, k^2_i,k_xk_z $   &  $b, k^2_i,k_xk_y$      & $(b+ a_ik^2_i ) $ \\ \hline \hline
$ d_z $   & $ a_0, k^2_i, k_yk_z$  &  $ k_yk_x, k_yk_z $  &  $ k_zk_x, k_zk_y $    & $c k_yk_z$          \\
\hline \hline
\end{tabular}
}
\caption{Mirror and ${\mathcal M}^{\prime}$ symmetry constraints on the bilinear crystalline momentum terms appearing in the low-energy two-band Hamiltonian ${\mathcal H}({\bf k})=\vec{\bf d}({\bf k}) \cdot \vec{\boldsymbol \sigma}$ assuming a magnetic field in the $\hat{y}$ direction. 
The last column indicates the bilinears compatible with the three symmetries. 
Additional rotational symmetries impose constraints on coefficients. For instance the ${\mathcal C}_{4y}$ symmetries implies that 
$a=c$ and also $a_x=a_z$. }
\label{table:symmetry}
\end{table}

Let us now add a Zeeman coupling, which, without loss of generality, we take due to a magnetic field along a principal crystallographic direction, which we dub  ${\hat y}$. This Zeeman coupling should be encoded in the usual term ${\mathcal H}_z=B_y \sigma_y$, which breaks the twofold degeneracy of the bands but does not yield band geometric properties. The crux of the story is that additional momentum-dependent terms can be activated by the Zeeman coupling. To show this, it is enough to note that with an external  magnetic field along a principal crystallographic direction the magnetic point group of the system is the centrosymmetric $4/mm'm'$. This group is generated by a residual mirror symmetry ${\mathcal M}_y$, a fourfold rotation with the rotation axis along the applied field, and two combined antiunitary symmetries ${\mathcal M}_{x,z} \Theta$ squaring to one. Using the transformation properties of Pauli matrices and momentum bilinears under the mirror and the antiunitary ${\mathcal M}^{\prime}$ symmetries listed in Table~\ref{table:symmetry}, we find that the breaking of time-reversal and point-group symmetries due to the application of the external magnetic field yields a generalized Zeeman  Hamiltonian with momentum dependent terms 
\begin{equation}
{\mathcal H}_z=a  k_xk_y \sigma_x + b \sigma_y + a k_yk_z  \sigma_z,
\label{eq:zeemangeneralized}
\end{equation}
where we have neglected the subleading terms (see Table~\ref{table:symmetry}) of the form $k_{x,y,z}^2 \sigma_y$. 
The coupling constants $a,b$ must be odd functions of the magnetic field strength, which implies that at weak fields they will grow linearly with $B_y$.

\begin{figure}[t] 
\centering
\resizebox{\columnwidth}{!} {    
\includegraphics[width=0.5\columnwidth]{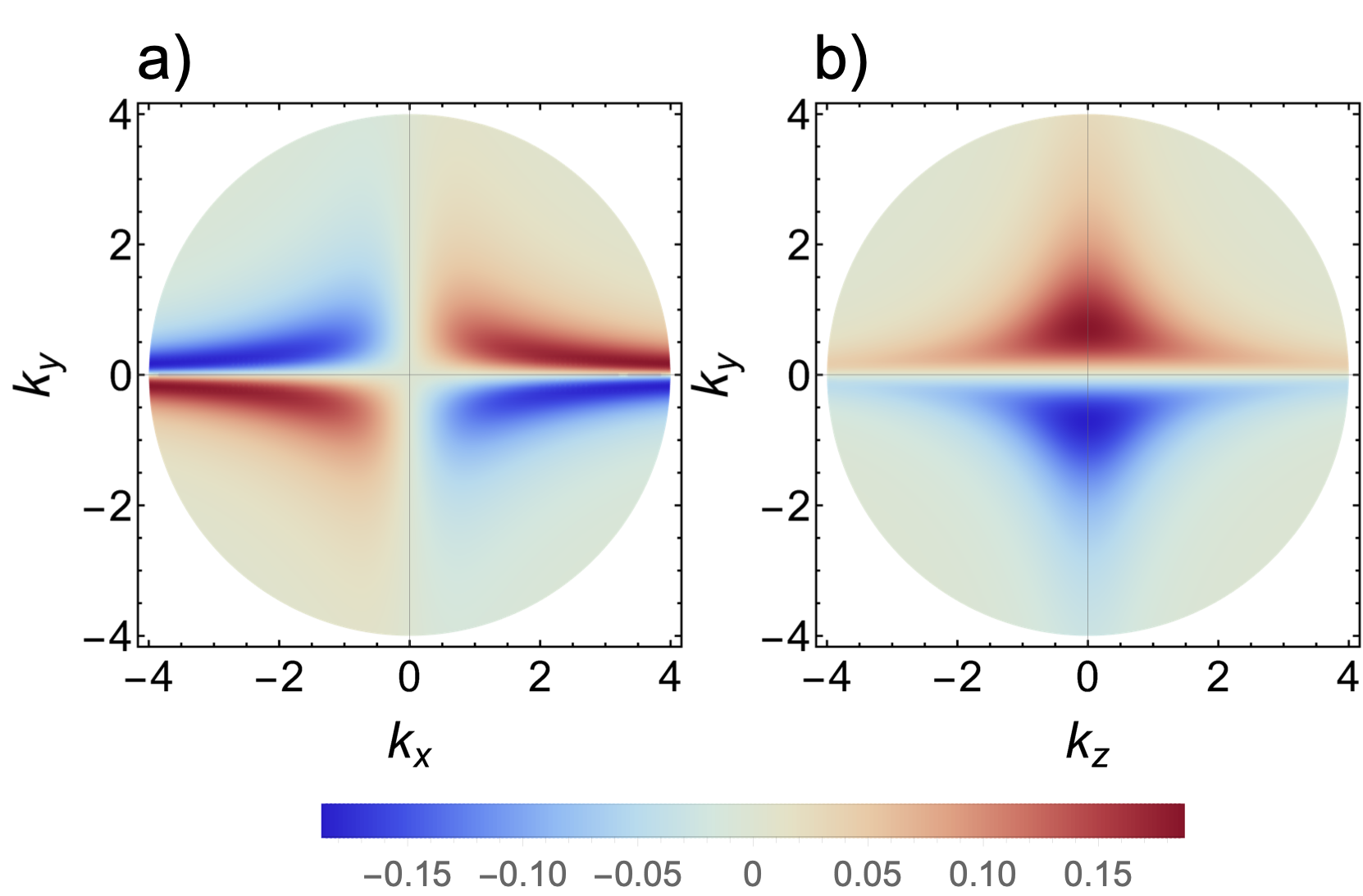}}
\caption{
Density plots of the $d$-wave Berry curvature $\Omega_x^{+}({\bf k})$ induced by the generalized Zeeman coupling in the $k_x,k_y$ and $k_y,k_z$ planes. In both cases we have chosen $b=a=1$.}
\label{fig:patternBC}
\end{figure}

The generalized Zeeman Hamiltonian written above leads  to an anisotropy in the energy spectrum that is directly proportional to the magnetic field strength. Perhaps more importantly, the electronic states are now equipped with a $d$-wave BC [see Fig.~\ref{fig:patternBC}], which can be explicitly written as 
\begin{equation}
\label{eq:BCvector}
\vec{\boldsymbol \Omega}^{\pm}({\bm k})= 
\mp    \frac{a^2 b}{2 [b^2+a^2 k^2_y(k^2_x+k^2_z) ]^{3/2}} \left\{  k_x k_y ,  -k^2_y , k_y k_z \right\}.
\end{equation}
In the equation above, the $\pm$ sign distinguishes the two spin-split bands forming the Kramers' pair. 
A few remarks are in order here. First, the BC written above scales as $B_y/|B_y|=\textrm{sign}(B_y)$ in the weak-field limit. The BC of a single band displays a discontinuity as $B_y \rightarrow 0$, which guarantees that precisely at $B_y \equiv 0$, {\it i.e.} in presence of time-reversal and inversion symmetries, the BC vanishes. 
Second, the fact that the BC of the two bands is opposite implies that any net anomalous Hall contribution, if existing, can only come from the difference in the shape of the Fermi surfaces that instead scales linearly with $B_y$. 
Third, in spin-orbit free systems any BC is precluded: the $SU(2)$ spin symmetry at $B_y\equiv 0$ implies that the generalized Zeeman coupling of Eq.~\ref{eq:zeemangeneralized} can be expressed using a single Pauli matrix even if momentum dependent terms are allowed.

Before discussing the consequences that Eq.~\ref{eq:BCvector} entails in magnetotransport, we show how the generalized Zeeman coupling of Eq.~\ref{eq:zeemangeneralized} can arise in a cubic system. Consider a time-reversal symmetric four-band model for two Kramers doublets of bands (one conduction and one valence) of opposite parity -- for instance deriving from an atomic $s$-orbital and a $p_{1/2}$ atomic orbital. The criterion derived by Fu and Kane~\cite{Fu2007} asserts that the level ordering between the two pairs of bands distinguishes a normal from a non-trivial topological class. Crucially,  the occurrence of the Zeeman-activated BC and its consequences on magnetotransport are independent of whether the two pairs undergo level crossing  and hence on the topology of the electronic states. 
The effective ${\bf k} \cdot {\bf p}$ Hamiltonian away from the $\Gamma$ point can be obtained in a straightforward manner. Since all the states have definite parity eigenvalues and the momentum ${\bf p}$ has odd parity, only matrix elements between states of opposite parity are non-vanishing. This also implies that the effective Hamiltonian to the linear order in ${\bf k}$ corresponds to a three-dimensional Dirac theory similar to the one~\cite{Zhang2009,Liu2010} governing the electronic states in the bulk of Bi$_2$Se$_3$. 
By additional considering terms quadratic in ${\bf k}$ we obtain an effective Dirac Hamiltonian that can be written as 
\begin{equation}
{\mathcal H}_{\textrm{eff}}= \epsilon_0({\bf k}) \tau_0  \sigma_0 + {\mathcal M}({\bf k}) \tau_z \otimes \sigma_0 - \dfrac{P}{\sqrt{3}} \tau_x \otimes {\boldsymbol \sigma} \cdot {\bf k}, 
\end{equation}
where the $\boldsymbol \sigma$ and $\boldsymbol \tau$ matrices act in spin and orbital space respectively, $\epsilon_0({\bf k})= \bar{M}  k^2 $ is a particle-hole symmetry breaking term, and ${\mathcal M}({\bf k})=M_0 + M_1 k^2$. The relative sign between $M_0$ and $M_1$ governs the band ordering~\cite{Liu2010} and thus distinguishes a normal insulator from a strong three-dimensional topological insulator. The parameter $P$ instead regulates the degree of non-parabolicity of the twofold degenerate electronic bands $\epsilon({\bf k})=\epsilon_0({\bf k}) \pm \left[ {\mathcal M}({\bf k})^2 + P^2 k^2 /3 \right]^{1/2}$.
We emphasize that in the normal band ordering the model written above  describes the $\Gamma_6$ $s$-bands and the spin-orbit split-off $\Gamma_7$ bands in cubic semiconductors.  
We next add a Zeeman coupling with orbital dependent Land\'e $g$-factors due to a $B_y$ magnetic field, which can be thus  cast as 
\begin{equation}
\Delta {\mathcal H}_{\textrm{eff}}=\bar{g}~\mu_B~B_y \tau_0 \otimes \sigma_y + \delta g~\mu_B~B_{y} \tau_z \otimes \sigma_y
\end{equation}
where $\bar{g}$ and $\delta g$ are the average and the difference between the two orbital  Land\'e $g$ factor, respectively, and $\mu_B$ is the Bohr magneton.
By performing a L\"owdin partioning~\cite{RolandWinklerBook}  we find [see the Supplemental Material] that in, for instance,  the conventional band ordering and for the pair of conduction bands, the projected two-band Hamiltonian has a generalized Zeeman coupling precisely of the Eq.~\ref{eq:zeemangeneralized} form with the two couplings $a=\left(\bar{g}-\delta g\right) \mu_B P^2 B_y/(6 M_0^2)$ and $b=\left(\bar{g}+\delta g\right)~\mu_B~B_y$. The linear $B_y$ dependence of these couplings guarantees that the ensuing BC is independent of the magnetic field strength. 
Furthermore, the BC is only dependent on the ratio between the two orbital dependent Land\'e $g$ factors and not on their absolute values. 
We have also computed the BC of the four-band model Hamiltonian ${\mathcal H}_{\textrm{eff}} + \Delta {\mathcal H}_{\textrm{eff}}$ using the method outlined in Ref.~\cite{GrafAnsgar}. In perfect agreement with our foregoing analysis, we find [see the Supplemental Material] the existence of weak field contributions, opposite for the two bands, scaling as $B_y/|B_y|$ and with the same crystalline momentum dependence of Eq.~\ref{eq:BCvector}, with one notable exception. The $\Omega_y({\bf k})$ component displays a momentum-independent contribution that is completely allowed, but it is not predicted by our symmetry-based theory. We will now show that remarkably the weak-field $\Omega_y({\bf k})$ BC does not contribute to the NLMR: the transport properties derived by the symmetry-based and the model Hamiltonian approaches are equivalent.


\begin{figure}[tbp]
\centering
\includegraphics[width=1\columnwidth]{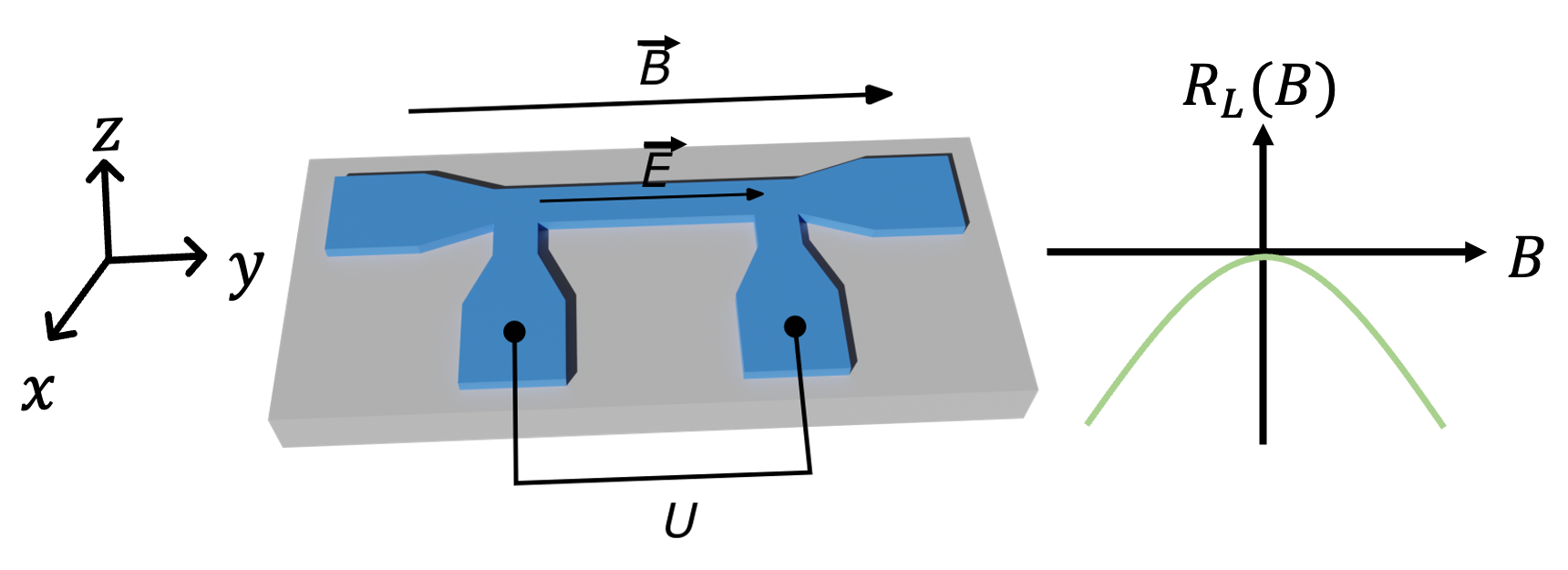}
\\ \vspace{0.6cm} 
\includegraphics[width=.9\columnwidth]{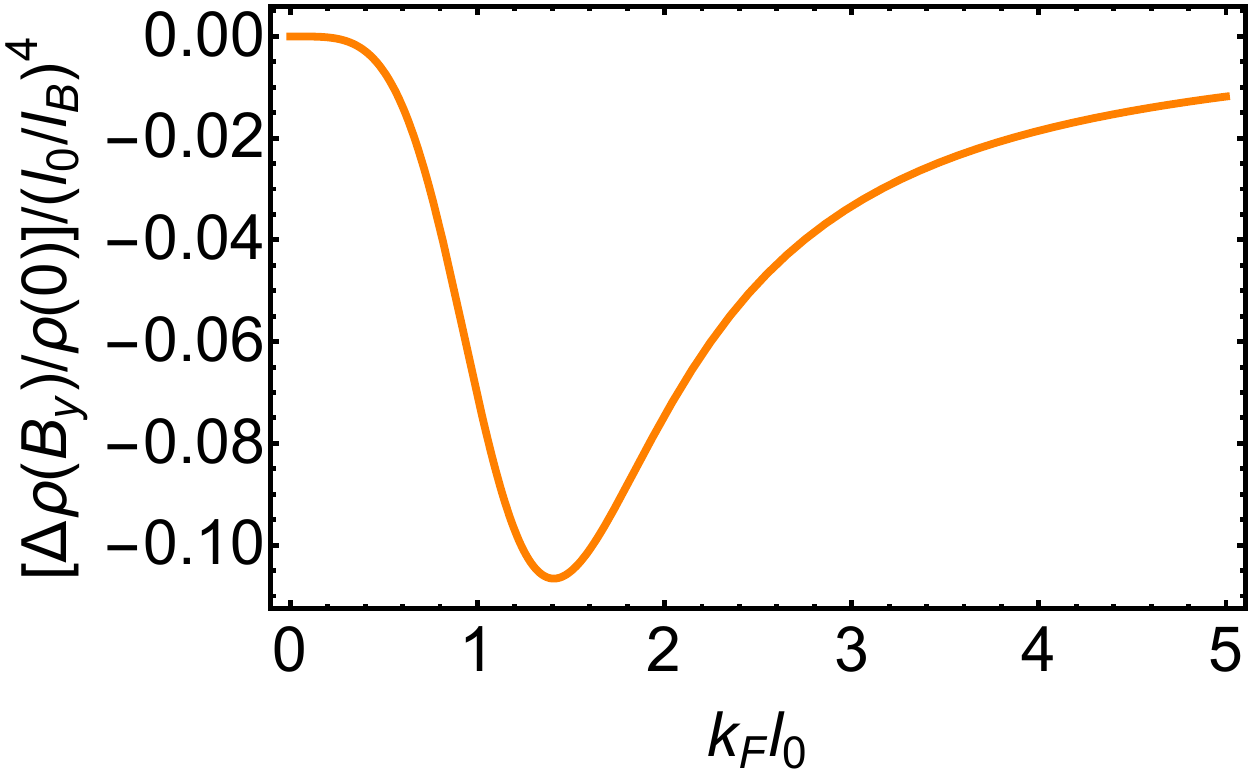}
\caption{Top pannel: Sketch of a device for longitudinal magnetoresistance measurement with the direction of the driving electric field and the applied in-plane magnetic field. The right panel shows an ensuing negative longitudinal magnetoresistance quadratic in the magnetic field. 
Bottom pannel: Behavior of the negative magnetoresistance $\Delta \rho(B_y)/\rho(0)$ obtained using the generalized Zeeman coupling Eq.~\ref{eq:zeemangeneralized} 
as a function of the Fermi wavevector $k_F$ measured in units of the inverse of the characteristic material dependent length scale $l_0=\sqrt{a/b}$. The magnetoconductivity is rescaled by the factor $(l_0/l_B)^4$ with $l_B=\sqrt{\hbar / (e B_y)}$ the Landau magnetic length.}
\label{fig:magnetoresistance}
\end{figure}

\paragraph{Semiclassical transport theory -- } The semiclassical equation of motion of charge carriers incorporating the BC of Bloch electrons~\cite{Sundaram1999,ChangMing2008,XiaoDi2010} are given by $\dot{\bm r}= (1/\hbar)\bm \nabla_{\bm k} \epsilon_{\bm k}  -\dot{\bm k}\times \bm \Omega$ and 
$\dot{\bm k}= (-e/\hbar) \left( \bm E + \dot{\bm r}\times \bm B \right)$. These coupled equations can be solved for $\dot{\bm r}$ and $\dot{\bm k}$ to obtain 

\begin{equation}
\label{Eq:velocity}
\dot{\bm r}=
D(\bm k)
\left[
\frac{1}{\hbar} \frac{\partial \epsilon_{\bm k}}{\partial \bm k} + \frac{e}{\hbar }\bm E \times \bm \Omega
+
\frac{e}{\hbar } \left( \frac{1}{\hbar} \frac{\partial \epsilon_{\bm k}}{\partial \bm k} \cdot \bm \Omega \right) \bm B
\right]
\end{equation}
\begin{equation}
\label{Eq:force}
\dot{\bm k}=
D(\bm k)
\left[
-\frac{e}{\hbar }\bm E - 
\frac{e}{\hbar^2 } \frac{\partial \epsilon_{\bm k}}{\partial \bm k} \times \bm B 
-\frac{e^2}{\hbar^2 } (\bm B \cdot \bm E)\bm \Omega
\right],
\end{equation}
with the factor $D(\bm k)=\left( 1+\frac{e}{\hbar} \bm B \cdot \bm \Omega \right)^{-1}$. The first and second terms in the r.h.s. of Eq.~\eqref{Eq:force} are the classical electric and Lorentz force respectively. 
The last term instead is caused by the BC of Bloch electrons and represents a force that is non-vanishing even for collinear $\bm E$ and $\bm B$. To proceed further, we use that the electronic distribution function $f_{\bm k}$ obeys the Boltzmann equation $\partial_t f_{\bm k}+\dot{\bm r} \cdot \bm \nabla_{\bm r}f_{\bm k}+\dot{\bm k} \cdot \bm \nabla_{\bm k}f_{\bm k} = \mathcal{I}_{coll}[f_{\bm k}]$, which we solve [see the Supplemental Material] in the relaxation time approximation where the collision integral $\mathcal{I}_{coll}[f] = - g_{\bm k} / \tau $ and $g_{\bm k}$ is the deviation of the electronic distribution function from the equilibrium $f_{eq}$ Fermi-Dirac distribution. By computing the electric current ${\bm j} = -e~{\rm tr}(\dot{\bm r} f_{\bm k})$, we subsequently obtain [see the Supplemental Material] the conductivity $\sigma_{ii}= \partial j_i/\partial E_i$. This can be written in a classical Drude form $\sigma_{ii}(\bm B)=e^2 \tau \sum_{\bf k} \dot{r}_{i} \dot{r}_{i} (-\partial f_{eq} / \partial \epsilon_{\bm k})$ with a BC-induced renormalization of the density of states~\cite{DiXiao2005} 
$\sum_{\bm k} =\int ( d^d\bm k/(2\pi)^d)  \left( 1+\frac{e}{\hbar} \bm B \cdot \bm \Omega \right)$, and the velocity components given by Eq.~\ref{Eq:velocity} apart from the anomalous velocity term governing the anomalous Hall conductivity~\cite{Haldane2004,Jungwirth2002,DimiCulcer2003,Nagaosa2006,Sinitsyn_2007JPCM,Nagaosa2010,ChenHua2014,BattilomoRaffaele2021,BurgosPRR2022}. Explicitly, we have 
\begin{align}
\sigma_{ii}(\bm B)
&=
e^2\tau 
\int \frac{ d^3\bm k }{ (2\pi)^3 }  
\frac{ \left[ v_i + \frac{e}{\hbar } \left(\bm v \cdot \bm \Omega \right) B_i \right]^2}
{\left( 1+\frac{e}{\hbar} \bm B \cdot \bm \Omega \right)}
\left(-\frac{\partial f_{eq}}{\partial \epsilon_{\bm k}} \right). 
\end{align}
where we introduced the group velocity $\bm v =\hbar^{-1} \bm \nabla_{\bm k}\epsilon_{\bm k}$. In the zero temperature limit $-\partial f_{eq} / \partial \epsilon_{\bm k} \approx \delta(\epsilon_F - \epsilon_{0\bm k})$, the zero-field conductivity corresponds to the usual Drude term $\sigma_{ii}(0)=e^2 \tau k_F^3 / (3 m^{\star} \pi^2)$. 
At finite ${\bm B}$, there are two field-induced corrections to the conductivity that must be taken into account. The first correction is due to the Zeeman-induced correction to the carrier group velocity whereas the second one is due to Zeeman-activated BC. Crucially, while the former is proportional to the Zeeman coupling strength (via the Bohr magneton $\mu_B$) the latter can only depend on the ratio between orbital-dependent effective Land\'e $g$-factors, and will thus yield the largest correction to the conductivity. Additionally, all corrections linear in magnetic fields vanish in agreement with the Onsager's relation $\sigma(\bm B)=\sigma(-\bm B)$. The leading correction to the Drude conductivity [see the Supplemental Material] can be finally written as  
\begin{align}
\label{Eq:MCBY}
\Delta \sigma_{yy}(B_y)
=
\dfrac{2 e^4\tau}{\hbar^2} B_y^2
\int \frac{d^3\bm k}{ (2\pi)^3} 
 (v_{x 0}\Omega_x+ v_{z 0}\Omega_z)^2
\left(-\frac{\partial f_{eq}}{\partial \epsilon_{0 \bm k}} \right), 
\end{align}
where $v_{x,z}^0$ and $\epsilon_{0 {\bf k}}$ indicate the bare group velocities and the band energy in the absence of the generalized Zeeman coupling. In deriving the equation above, we also used that
the contributions of the two conduction bands sum up due to the fact that only bilinears of the (opposite in sign) BC components appear. 
Eq.~\ref{Eq:MCBY} is the second  main result of this study: it
provides the expression for a magnetic field-induced increase of conductivity in a spin-orbit coupled material with a centrosymmetric cubic crystalline arrangement. The ensuing NLMR  $ \Delta \rho_{yy}(B_y)/\rho(0) \simeq - \Delta \sigma_{yy}(B_y)/\sigma(0) $ grows quadratically with the externally applied magnetic field strength [see Fig.~\ref{fig:magnetoresistance}] while being independent of the relaxation time and thus intrinsic. 

A direct computation of Eq.~\ref{Eq:MCBY} for the generalized Zeeman coupling in Eq.~\ref{eq:zeemangeneralized} leads to the NLMR shown in Fig.~\ref{fig:magnetoresistance}. It scales as $(l_0/l_B)^4 f(k_F~l_0)$  [see the Supplemental Material] where $l_B=\sqrt{\hbar/e B_y}$ is the magnetic length while $l_0=\sqrt{a/b}$ is a material-dependent characteristic length. The Fermi momentum function 
$f(k_F l_0)$ is instead independent of specific material properties and hence universal. 
As discussed before, the latter can be also expressed in terms of the linear in ${\bf k}$ coupling $P$ between Kramers' pairs of opposite parity, directly controlling the non-parabolicity of the bands, and the direct gap between conduction and valence bands $M_0$. Additionally, the NLMR displays a non-monotonous behavior that is a characteristic fingerprint of BC-mediated transport effects, with a maximum value $\Delta \rho_{yy}(B_y)/\rho(0) \simeq 0.1 \left(l_0/l_B\right)^4$. For Bi$_2$Se$_3$ we can estimate~\cite{Nechaev2016} $l_0 \simeq 2$nm, and thus a characteristic NLMR of the order of $0.5 \%$ at a few Teslas which is close to the values experimentally observed~\cite{Wiedmann2016}.
We note that this NLMR value is independent of the $g$-factor, in marked contrast with a previous theoretical estimate~\cite{DaiXin2017} where a large Land\'e $g$-factor of $30$ has been used.

\paragraph{Conclusions --} Thus far, NLMR in  centrosymmetric and time-reversal invariant conductors has been shown to occur either in 
the quantum regime $\omega_c \tau \gg 1$ with well-formed Landau levels ~\cite{Andreev2018,Goswami2015} or considering very large Land\'e $g$-factors as a result of the orbital magnetic moment of Zeeman-split electronic bands~\cite{DaiXin2017}. We have instead 
identified a previously unknown mechanism of NLMR at weak fields in centrosymmetric and time-reversal invariant materials. We have shown that the non-parabolicity of spin-orbit coupled bands leads to a generalized Zeeman coupling in the projected doublet of  bands participating in transport. This generalized Zeeman coupling contains momentum dependent terms that lead to a $d$-wave Berry curvature independent of the magnetic field strength, which directly couples to the orbital motion of Bloch electrons. In contrast to the mechanism based on Zeeman-split bands~\cite{DaiXin2017}, our theory is independent of the Land\'e $g$-factor and the orbital magnetic moment. 
The NLMR mechanism individuated in this study can appear in all centrosymmetric point groups [see the Supplemental Materials]. Additionally, it is independent of the topology of the electronic bands and can appear both in doped topological insulators and in conventional conductors. We therefore expect the weak-field NLMR to act full force in a vast class of non-magnetic spin-orbit coupled materials, including  for instance Bi$_{1-x}$Sb$_x$. In this compound, the electronic pockets at the $L$ point of the Brillouin zone  have a small direct energy gap and a strongly nonparabolic, nearly linear dispersion~\cite{Hasan2010} for small $x$. This can lead to strongly enhanced values of the characteristic length $l_0$ and consequently of its intrinsic weak-field NLMR. 

\begin{acknowledgments}
We acknowledge support by the Italian Ministry
of Foreign Affairs and International Cooperation, grant
PGR12351 “ULTRAQMAT”, the PNRR MUR project PE0000023-NQSTI, 
and the European Union ERC grant 3D multiFerro (Project number: 101141331).
\end{acknowledgments}

%


\newpage
\onecolumngrid

\begin{center}
    \textbf{ Supplemental Material for}\\[0.5em]
    \textbf{``Zeeman-activated Berry curvature  magnetotransport \\ from the bulk of  non-magnetic metals with inversion symmetry''}\\[1em]

    \text{Rhonald Burgos Atencia}$^{1}$, 
    \text{Antonio Vecchione}$^{2}$, 
    \text{Denys Makarov}$^{3}$, 
    \text{Carmine Ortix}$^{1,2}$\\[0.5em]

    {\small
    $^{1}$\textit{Dipartimento di Fisica "E. R. Caianiello", Università di Salerno, IT-84084 Fisciano (SA), Italy}\\
    $^{2}$\textit{CNR-SPIN, c/o Università di Salerno, IT-84084 Fisciano (SA), Italy}\\
    $^{3}$\textit{Helmholtz-Zentrum Dresden-Rossendorf e.V., Institute of Ion Beam Physics and Materials Research, 01328 Dresden, Germany}
    }
\end{center}

\vspace{1em}

\section{Generalized Zeeman coupling from the four-band model Hamiltonian}

Let us consider a four-band Hamiltonian for two Kramers' pairs of electronic bands in an ${\mathcal O}_h$ crystalline environment
\begin{equation}
{\mathcal H}_{\textrm{eff}}= \epsilon_0({\bf k}) \tau_0  \sigma_0 + {\mathcal M}({\bf k}) \tau_z \otimes \sigma_0 - \dfrac{P}{\sqrt{3}} \tau_x \otimes {\boldsymbol \sigma} \cdot {\bf k}.
\end{equation}

The eigenvalues are given by $\epsilon^{\pm}({\bf k})=\epsilon_0({\bf k}) \pm \left[ {\mathcal M}({\bf k})^2 + P^2 k^2 /3 \right]^{1/2}$ 
where ${\mathcal M}({\bf k})=M_0 + M_1 k^2$. Now we add a Zeeman coupling which is given by the expression 
\begin{equation}
\Delta {\mathcal H}_{\textrm{eff}}=\bar{g}~\mu_{B}~B_y \tau_0 \otimes \sigma_y + \delta g~\mu_{B}~B_{y} \tau_z \otimes \sigma_y. 
\end{equation}

In matrix form it reads
\begin{equation}
\Delta {\mathcal H}_{\textrm{eff}}
=
\begin{pmatrix} 
  0         & -i(\bar{g} + \delta g )~\mu_{B} B_y & 0 & 0 \\
i (\bar{g} + \delta g )~\mu_{B} B_y &    0        & 0 & 0 \\
      0        & 0 & 0 & -i(\bar{g} - \delta g )~\mu_{B}B_y \\     
  0 & 0 & i(\bar{g} - \delta g )~\mu_{B} B_y & 0 \\     
\end{pmatrix}.
\end{equation}

Let us consider the unperturbed Hamiltonian
\begin{equation}
H_0=
\begin{pmatrix} 
\epsilon_{0\bm k} +\mathcal{M}(\bm k)   & 0  & 0 & 0  \\
0   &  \epsilon_{0\bm k} +\mathcal{M}(\bm k) &  0 & 0  \\
0 & 0 & \epsilon_{0\bm k} - \mathcal{M}(\bm k)   & 0 \\
0   & 0  & 0  & \epsilon_{0\bm k} - \mathcal{M}(\bm k)  \\
\end{pmatrix}
\end{equation}
with the corresponding degenerate eigenvectors
\begin{equation}
|+ \uparrow \rangle_{1}=
\begin{pmatrix} 
1\\
0\\
0\\
0\\
\end{pmatrix} \quad
| + \downarrow \rangle_{2}=
\begin{pmatrix} 
0\\
1\\
0\\
0\\
\end{pmatrix} \quad
| - \uparrow \rangle_{3}=
\begin{pmatrix} 
0\\
0\\
1\\
0\\
\end{pmatrix} \quad
| - \downarrow \rangle_{4}=
\begin{pmatrix} 
0\\
0\\
0\\
1\\
\end{pmatrix}.
\end{equation}

The perturbation reads
\begin{equation}
H'=
\begin{pmatrix} 
                 0                  & -i(\bar{g} + \delta g) ~\mu_{B} B_y              & -\frac{1}{\sqrt{3}}Pk_{z}    & -\frac{1}{\sqrt{3}}Pk_{-}    \\
      i(\bar{g} + \delta g) ~\mu_{B}B_y        &  0                                  &  -\frac{1}{\sqrt{3}}Pk_{+}   & \frac{1}{\sqrt{3}}Pk_{z}   \\
-\frac{1}{\sqrt{3}}Pk_{z}  & -\frac{1}{\sqrt{3}}Pk_{-}  &                   0                    &  -i(\bar{g} - \delta g)~\mu_{B} B_y   \\
-\frac{1}{\sqrt{3}}Pk_{+}  & \frac{1}{\sqrt{3}}Pk_{z}  &             i(\bar{g} - \delta g)~\mu_{B} B_y              &   0  \\
\end{pmatrix}. 
\end{equation}

The Hamiltonian is now understood as $H=H_0 + H'$. In the following we will apply the L\"owdin partioning~\cite{RolandWinklerBookSM} method in order to find an effective
$2\times 2$ effective Hamiltonian.

\subsection{The L\"owdin partioning  up to the third order}

Note that the Zeeman term is already in the block diagonal sector. This means that to the first order in perturbation theory, the effective term is 
\begin{equation}
[H^{(1)}]=
(\bar{g}+\delta g) ~\mu_{B}B_y 
\begin{pmatrix} 
                    0                & -i       \\
      i    &         0                    \\
\end{pmatrix} =
(\bar{g}+\delta g)~\mu_{B}B_y \sigma_y. 
\end{equation}

\subsubsection{Second order terms}

To the second order in perturbation theory we need to compute the quantity 
\begin{equation}
[H^{(2)}]_{mm'}=
\frac{1}{2}\sum_{l\neq (m,m')}[H']_{ml} [H']_{lm'}\left[ \frac{1}{E_{m}-E_{l}} + \frac{1}{E_{m'}-E_{l}} \right]
\end{equation}
where $(m,m')$ belong to the upper sub-space, namely they take values  $m,m'=1,2 $ while $l$ belongs to the bottom sub-space namely $l= 3,4$. Then
we have that $E_{m}-E_{l}=E_{m'}-E_{l}= 2\mathcal{M}(\bm k)$. Second order perturbation theory then gives
\begin{align}
[H^{(2)}]_{11} 
=\frac{1}{3}\frac{P^2k^2}{2\mathcal{M}(\bm k)} = [H^{(2)}]_{22}.
\end{align}

The off-diagonal elements are zero, namely, $[H^{(2)}]_{12}=[H^{(2)}]_{21}=0$. 
Then, the second order perturbation theory gives the effective Hamiltonian 
\begin{equation}
[H^{(2)}]=
\frac{1}{ 2\mathcal{M}(\bm k) }\frac{P^2k^2}{ 3 } 
\begin{pmatrix} 
                    1                 &  0     \\
       0     &          1                      \\
\end{pmatrix}.
\end{equation}

This is a renormalized parabola.

\subsubsection{Third order terms}

Up to third order perturbation theory we need to compute the quantities
\begin{align}
[H^{(3)}]_{mm'}
&=-\frac{1}{2}\sum_{l}\sum_{m''}
\left[
\frac{[H']_{ml}[H']_{lm''}[H']_{m''m'} }{ ( E_{m'} - E_{l})(E_{m''}-E_{l}) }
\right] \\
&-\frac{1}{2}\sum_{l}\sum_{m''}
\left[
\frac{ [H']_{mm''}[H']_{m''l}[H']_{lm'} }{(E_m - E_l )( E_{m''} - E_l ) } 
\right] \\
&+\frac{1}{2}\sum_{l,l'}
[H']_{ml}[H']_{ll'}[H']_{l'm'}
\left[ \frac{1}{ (E_m - E_l )(E_m - E_{l'}) } + \frac{1}{ (E_{m'} - E_l )(E_{m'} - E_{l'} ) }  \right].
\end{align}

The full off-diagonal term reads 
\begin{align}
[H^{(3)}]_{12}
&=
\frac{1}{2} \frac{( P/\sqrt{3})^2 }{(\mathcal{M}(\bm k))^2} (\bar{g} - \delta g) ~\mu_{B}  k_xk_yB_y  
+
i \frac{1}{2} \frac{ ( P/\sqrt{3})^2 }{ (\mathcal{M}(\bm k))^2 } 
[\bar{g}~\mu_{B}  B_y(k^2_x+k^2_z) +  \delta g~\mu_{B}  B_y k^2_y].
\end{align}

In the same manner we calculate the diagonal terms of the third order perturbation theory. We get
\begin{align}
[H^{(3)}]_{11}=\frac{1}{2} \frac{ ( P/\sqrt{3})^2 }{ (\mathcal{M}(\bm k))^2 }( \bar{g} - \delta g)~\mu_{B} B_y k_y k_z  
\end{align}
and $[H^{(3)}]_{22}=-[H^{(3)}]_{11}$. With the first order perturbation result 
we can define the effective Hamiltonian of the form $[H^{(1)}]+[H^{(3)}]= \bm d \cdot \bm \sigma$ where
\begin{align}
d_x&= \frac{1}{3}\frac{ P^2 }{(2\mathcal{M}(\bm k))^2}   (\bar{g} - \delta g)~\mu_{B} 2 B_y k_xk_y \\
d_y&= ( \bar{g} + \delta g)~\mu_{B}B_y -\frac{1}{3}\frac{ P^2 }{ (2\mathcal{M}(\bm k))^2 }  [ 2\bar{g}~\mu_{B} B_y(k^2_x+k^2_z) +2\delta g~\mu_{B} B_y k^2_y    ]\\
d_z&= \frac{1}{3}\frac{ P^2 }{ (2\mathcal{M}(\bm k))^2 }  (\bar{g} - \delta g) ~\mu_{B} 2B_y k_yk_z,
\end{align}
which to the leading order in momentum can be expressed as 
\begin{align}
\label{eq:effectivevector}
d_x& \approx \frac{1}{3}\frac{ P^2 }{ 2 M^2_0}  B_y (\bar{g} -\delta g)~\mu_{B} k_xk_y \\
d_y& \approx ( \bar{g} + \delta g)~\mu_{B} B_y 
 \\
d_z& \approx \frac{1}{3}\frac{ P^2 }{ 2M^2_0 } (\bar{g} -\delta g)~\mu_{B} B_y k_yk_z.
\end{align}

The effective $2\times 2$ Hamiltonian will be $H=H_0+H^{(1)} + H^{(2)} + H^{(3)}$.
\begin{equation}
H_0= [ \epsilon_{0\bm k} +\mathcal{M}(\bm k)]\sigma_0 + \frac{1}{ 2\mathcal{M}(\bm k) }\frac{P^2k^2}{ 3 } \sigma_0+\bm d \cdot \bm \sigma.
\end{equation}

\subsection{Berry curvature in the four-band model}

In this section we show that the actual $4\times 4$ Hamiltonian hosts a Berry curvature independent of the external in-plane magnetic field. 
We follow the procedure described in Ref~\cite{GrafAnsgarSM}.
The Hamiltonian can be written as $H=h_0\lambda_0+ \bm h \cdot \bm \lambda$ where $\lambda_0$ is the identity matrix and  $\bm \lambda$ are 15 generalized Gell-mann matrices given by 
\begin{align}
\lambda_1 &= 
\begin{pmatrix}
0 & 1 & 0 & 0 \\
1 & 0 & 0 & 0 \\
0 & 0 & 0 & 0 \\
0 & 0 & 0 & 0
\end{pmatrix}
\quad
\lambda_2 = 
\begin{pmatrix}
0 & -i & 0 & 0 \\
i & 0 & 0 & 0 \\
0 & 0 & 0 & 0 \\
0 & 0 & 0 & 0
\end{pmatrix}
\quad
\lambda_3 = 
\begin{pmatrix}
1 & 0 & 0 & 0 \\
0 & -1 & 0 & 0 \\
0 & 0 & 0 & 0 \\
0 & 0 & 0 & 0
\end{pmatrix}
\quad
\lambda_4 = 
\begin{pmatrix}
0 & 0 & 1 & 0 \\
0 & 0 & 0 & 0 \\
1 & 0 & 0 & 0 \\
0 & 0 & 0 & 0
\end{pmatrix} 
\quad
\lambda_5 = 
\begin{pmatrix}
0 & 0 & -i & 0 \\
0 & 0 & 0 & 0 \\
i & 0 & 0 & 0 \\
0 & 0 & 0 & 0
\end{pmatrix}  \nonumber \\ 
\lambda_6 &= 
\begin{pmatrix}
0 & 0 & 0 & 0 \\
0 & 0 & 1 & 0 \\
0 & 1 & 0 & 0 \\
0 & 0 & 0 & 0
\end{pmatrix} 
\quad
\lambda_7 = 
\begin{pmatrix}
0 & 0 & 0 & 0 \\
0 & 0 & -i & 0 \\
0 & i & 0 & 0 \\
0 & 0 & 0 & 0
\end{pmatrix}
\quad
\lambda_8 = \frac{1}{\sqrt{3}}
\begin{pmatrix}
1 & 0 & 0 & 0 \\
0 & 1 & 0 & 0 \\
0 & 0 & -2 & 0 \\
0 & 0 & 0 & 0
\end{pmatrix}
\quad
\lambda_9 = 
\begin{pmatrix}
0 & 0 & 0 & 1 \\
0 & 0 & 0 & 0 \\
0 & 0 & 0 & 0 \\
1 & 0 & 0 & 0
\end{pmatrix}
\quad
\lambda_{10} = 
\begin{pmatrix}
0 & 0 & 0 & -i \\
0 & 0 & 0 & 0 \\
0 & 0 & 0 & 0 \\
i & 0 & 0 & 0
\end{pmatrix} \nonumber \\ 
\lambda_{11} &= 
\begin{pmatrix}
0 & 0 & 0 & 0 \\
0 & 0 & 0 & 1 \\
0 & 0 & 0 & 0 \\
0 & 1 & 0 & 0
\end{pmatrix}
\quad
\lambda_{12} = 
\begin{pmatrix}
0 & 0 & 0 & 0 \\
0 & 0 & 0 & -i \\
0 & 0 & 0 & 0 \\
0 & i & 0 & 0
\end{pmatrix} 
\quad
\lambda_{13} = 
\begin{pmatrix}
0 & 0 & 0 & 0 \\
0 & 0 & 0 & 0 \\
0 & 0 & 0 & 1 \\
0 & 0 & 1 & 0
\end{pmatrix}
\quad
\lambda_{14} = 
\begin{pmatrix}
0 & 0 & 0 & 0 \\
0 & 0 & 0 & 0 \\
0 & 0 & 0 & -i \\
0 & 0 & i & 0
\end{pmatrix}
\quad
\lambda_{15} = 
\frac{1}{\sqrt{6}}
\begin{pmatrix}
1 & 0 & 0 & 0 \\
0 & 1 & 0 & 0 \\
0 & 0 & 1 & 0 \\
0 & 0 & 0 & -3
\end{pmatrix}.
\end{align}

The vector Hamiltonian $\bm h$ has components $h_2=(\bar{g} + \delta g)~\mu_{B}B_y$, $h_4=-\frac{1}{\sqrt{3}}Pk_z$,
$h_6=-\frac{1}{\sqrt{3}}Pk_x$, $h_7=-\frac{1}{\sqrt{3}}Pk_y$, 
$h_8=\frac{\sqrt{3}}{3}\mathcal{M}(\bm k) $, 
$h_9=-\frac{1}{\sqrt{3}}Pk_x$,
$h_{10}=-\frac{1}{\sqrt{3}}Pk_y$, 
$h_{11}=\frac{1}{\sqrt{3}}Pk_z$, 
$h_{14}= (\bar{g} - \delta g)~\mu_{B} B_y$ and 
$h_{15}=\frac{\sqrt{6}}{6} \mathcal{M}(\bm k) $. Also $h_1=h_3=h_5=h_{12}=h_{13}=0$.
The component $h_0=\epsilon_{0\bm k}$, but it is not relevant for the calculation of the Berry curvature. 
The Berry curvature is given by 
\begin{equation}
\label{ }
\bm \Omega^{ij}_{\alpha}=-\frac{1}{2}\bm b_{\alpha} \cdot (\bm b^{i}_{\alpha} \times \bm b^{j}_{\alpha}),
\end{equation}
where $\bm b_{\alpha}$ are generalized Bloch vectors for the $\alpha$-band, $\bm b^{i}_{\alpha}=\partial \bm b_{\alpha}/\partial k_i$ and the vector product between the generalized Bloch 
vectors is defined as $(\bm m \times \bm n)_{a}=f_{abc}m_bn_c$ with the totally antisymmetric structure constant
$f_{abc} =(-i/4){\rm Tr}( [\lambda_a,\lambda_b]\lambda_c) $. The generalized Bloch vector follows as:
\begin{equation}
\label{ }
\bm b_{\alpha}=\frac{2}{4E^3_{\alpha} - C_2E_{\alpha} - C_3/3 }
\left[ \left( E^2_{\alpha} - \frac{C_2}{4} \right) \bm h + E_{\alpha}\bm h_{\star}  + \bm h_{\star \star} \right],
\end{equation}
where $\bm h_{\star} \equiv \bm h \star \bm h$ and $\bm h_{\star \star} \equiv \bm h \star \bm h_{\star}$. 
The $\star$-product is defined as $(\bm m \star \bm n)_a=d_{abc}m_bn_c$ with the totally symmetric structure constant $d_{abc}=(1/4){\rm Tr}[\{\lambda_a,\lambda_b\} \lambda_c]$. The Casimir invariants $C_{i}$ are $C_2=2|\bm h|^2$ and $C_3=2\bm h \cdot \bm h_{\star}$. Also, 
$E_{\alpha}$ are the eigenvalues of the Hamiltonian. In order to find close analytical expressions we take $\delta g=0$. The degeneracy is lifted and 
the energies for the two conduction bands are
\begin{equation}
E^{c}_{\pm} = \sqrt{M^2_0+ k^2( P/\sqrt{3})^2 + \delta g^2~\mu_{B}^2 B^2_y \pm 2 \delta |g~\mu_{B}| |B_y| \sqrt{M^2_0 + k^2_y(P/\sqrt{3})^2 } }
\end{equation}
and for the two valence bands are 
\begin{equation}
E^{v}_{\pm} = - \sqrt{M^2_0+ k^2( P/\sqrt{3})^2 + \delta g^2 ~\mu_{B}^2 B^2_y \pm 2 \delta |g~\mu_{B}| |B_y| \sqrt{M^2_0 + k^2_y(P/\sqrt{3})^2 } }.
\end{equation}

\begin{figure}[tbp] 
 \centering
\includegraphics[width=0.6\columnwidth]{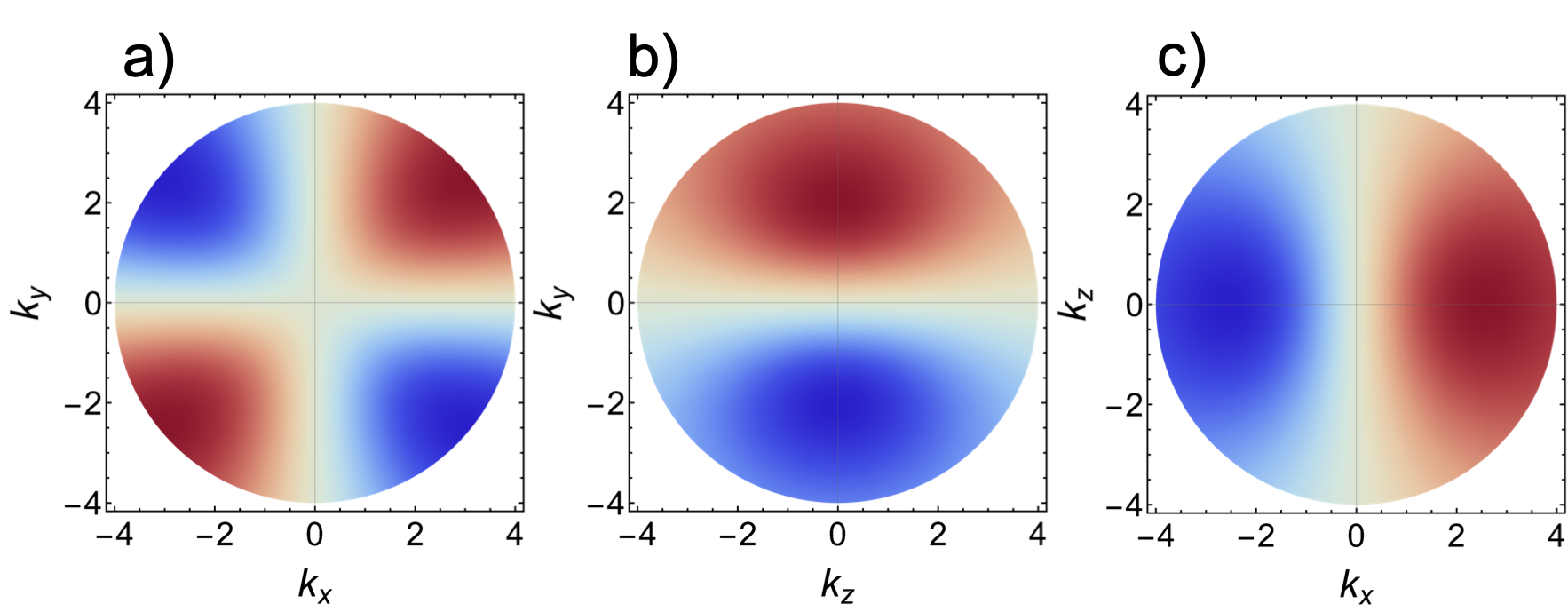} \\
\includegraphics[width=0.6\columnwidth]{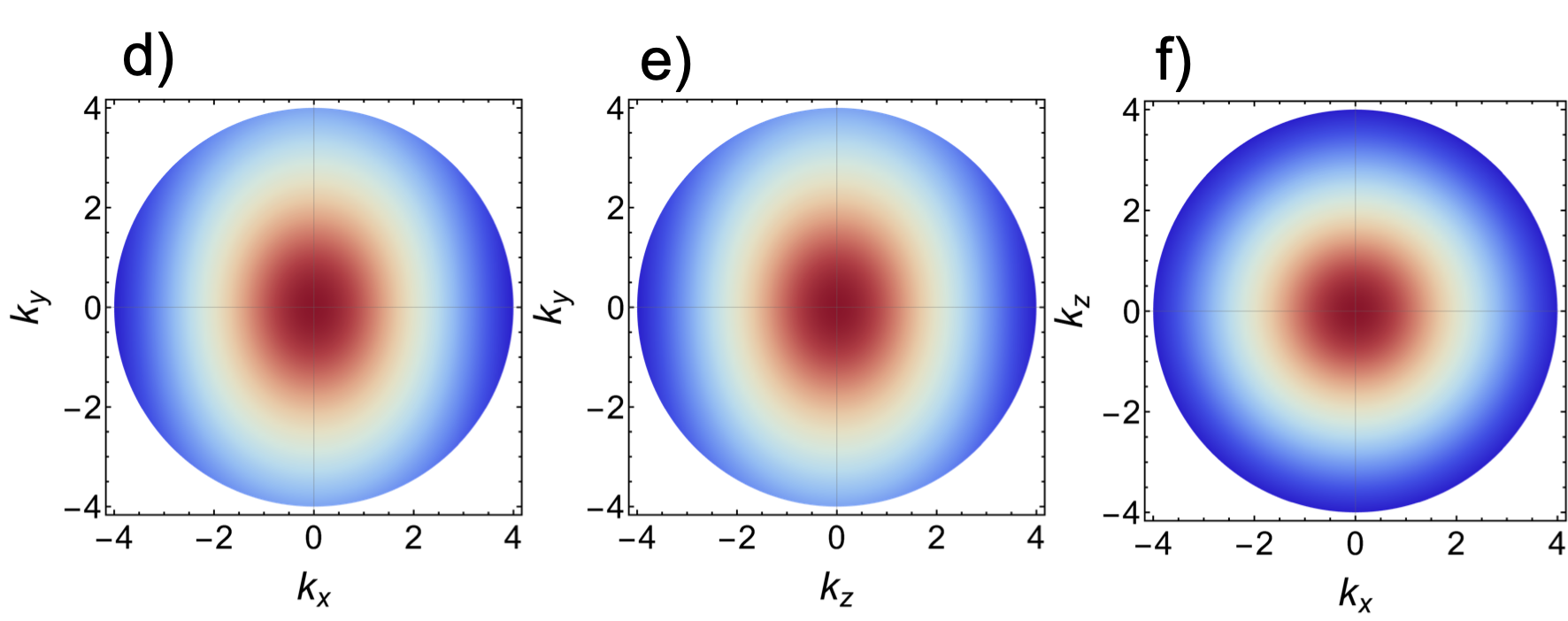} \\
\includegraphics[width=0.6\columnwidth]{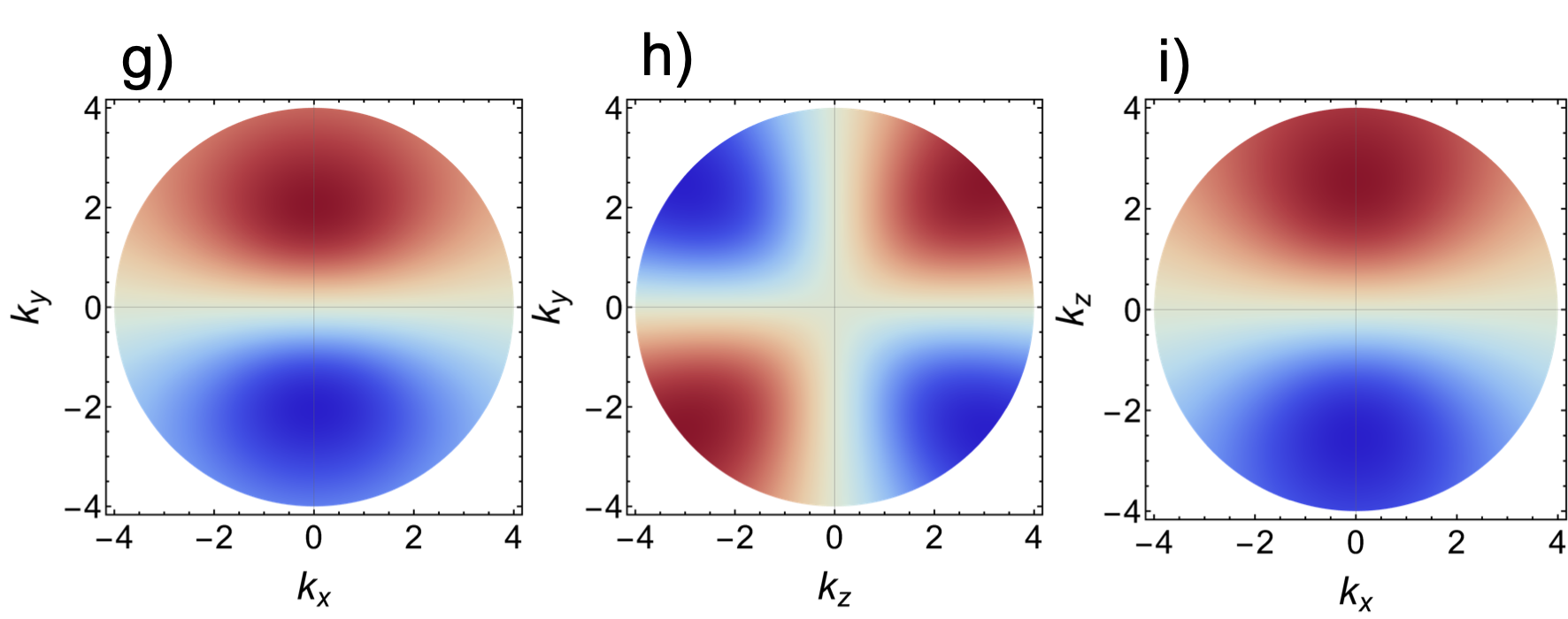}
\caption{Berry curvature for the four-band model. The upper panel shows a density plot of $\Omega^{+}_{\bm k, x}$ for the conduction band 
for different projection planes. The middle panel shows $\Omega^{+}_{\bm k, y}$ and the bottom panel shows $\Omega^{+}_{\bm k, z}$.}
 \label{fig:BC44all}
 \end{figure}

The Berry curvature vector for the two conduction bands has components  
\begin{align}
\Omega^{\pm  }_{\bm k, x}=\mp 
sign(B_y)
\frac{ k_xk_y (P/\sqrt{3})^4 }
{ 2[M^2_0 + k^2_y  (P/\sqrt{3})^2 ]^{1/2} [ M^2_0 + k^2 (P/\sqrt{3})^2]^{3/2}} 
+\mathcal{O}(B_y)
\end{align}
\begin{align}
\Omega^{\pm }_{\bm k, y}=\mp
sign(B_y)
\frac{ (P/\sqrt{3})^2 [M^2_0 + k^2_y  (P/\sqrt{3})^2 ]^{1/2} }
{ 2 [ M^2_0 + k^2 (P/\sqrt{3})^2]^{3/2}} 
+\mathcal{O}(B_y)
\end{align}
\begin{align}
\Omega^{\pm}_{\bm k, z}=\mp
sign(B_y)
\frac{ k_zk_y (P/\sqrt{3})^4 }
{2 [M^2_0 + k^2_y  (P/\sqrt{3})^2 ]^{1/2} [ M^2_0 + k^2 (P/\sqrt{3})^2]^{3/2}} 
+\mathcal{O}(B_y)
\end{align}

We present density plots in Fig.\eqref{fig:BC44all}.

\section{Semi-classical transport theory}

The Boltzmann equation for the distribution $f_{\bm k}$ reads
\begin{equation}
\frac{\partial f}{\partial t}+\dot{\bm r} \cdot \frac{\partial f}{\partial \bm r} + \dot{\bm k} \cdot \frac{\partial f}{\partial \bm k} = \mathcal{I}_{coll}[f]. 
\end{equation}

In the above equation we need the semiclassical velocity, namely 
\begin{equation}
\dot{\bm r}=\frac{1}{\hbar} \frac{\partial \varepsilon_{\bm k}}{\partial \bm k} -\dot{\bm k}\times \bm \Omega
\end{equation}
and the semiclassical variation of momentum, namely 
\begin{equation}
\label{Eq:velocity}
\dot{\bm k}=-\frac{e}{\hbar} \left[ \bm E + \dot{\bm r}\times \bm B \right].
\end{equation}

Solving the above coupled equations for $\bm r$ and $\bm k$ we get 
\begin{equation}
\dot{\bm r}=
\left( 1+\frac{e}{\hbar} \bm B \cdot \bm \Omega \right)^{-1}
\left[
\frac{1}{\hbar} \frac{\partial \varepsilon_{\bm k}}{\partial \bm k} + \frac{e}{\hbar }\bm E \times \bm \Omega
+
\frac{e}{\hbar } \left( \frac{1}{\hbar} \frac{\partial \varepsilon_{\bm k}}{\partial \bm k} \cdot \bm \Omega \right) \bm B
\right]
\end{equation}
\begin{equation}
\dot{\bm k}=
\left( 1+\frac{e}{\hbar} \bm B \cdot \bm \Omega \right)^{-1}
\left[
-\frac{e}{\hbar }\bm E - 
\frac{e}{\hbar } \left( \frac{1}{\hbar} \frac{\partial \varepsilon_{\bm k}}{\partial \bm k} \times \bm B \right)
-\frac{e^2}{\hbar^2 } (\bm B \cdot \bm E)\bm \Omega
\right].
\end{equation}

In the relaxation time approximation $\mathcal{I}_{coll}[f] = - g_{\bm k}/\tau$. In stationary states, the Boltzmann equation is 
$ \dot{\bm k} \cdot \frac{\partial f}{\partial \bm k} = - g_{\bm k}/\tau $.
We take $f=f_0+ g_{\bm k}$ where $f_{0}$ is the equilibrium distribution and  $g_{\bm k}$ is the deviation from equilibrium due to the electric field. 
The equation then reads
\begin{equation}
D\left[
-\frac{e}{\hbar }\bm E - 
\frac{e}{\hbar } \left( \frac{1}{\hbar} \frac{\partial \varepsilon_{\bm k}}{\partial \bm k} \times \bm B \right)
-\frac{e^2}{\hbar^2 } (\bm B \cdot \bm E)\bm \Omega
\right]
 \cdot 
\left( \frac{\partial f_0}{\partial \bm k} +  \frac{\partial g_{\bm k}}{\partial \bm k}\right) = - \frac{g_{\bm k}}{\tau } 
\end{equation}
with $D=(1+\frac{e}{\hbar} \bm B \cdot \bm \Omega)^{-1}$. In linear response
\begin{equation}
\label{Eq:BE}
D\left[
-\frac{e}{\hbar }\bm E
-\frac{e^2}{\hbar^2 } (\bm B \cdot \bm E)\bm \Omega
\right] 
\cdot 
\left( \frac{\partial \varepsilon_{\bm k}}{\partial \bm k} \right) 
\left( \frac{\partial f_0}{\partial \varepsilon_{\bm k}} \right)
+
D\left[
- \frac{e}{\hbar } \left( \frac{1}{\hbar} \frac{\partial \varepsilon_{\bm k}}{\partial \bm k} \times \bm B \right)
\right]
 \cdot 
\left(   \frac{\partial g_{\bm k}}{\partial \bm k}\right) 
= - \frac{g_{\bm k}}{\tau }.
\end{equation}

In terms of the group velocity we can write 
\begin{equation}
\label{}
D\left[
-e\bm E\cdot \bm v
-\frac{e^2}{\hbar } (\bm B \cdot \bm E) (\bm \Omega \cdot \bm v )
\right] 
\left( \frac{\partial f_0}{\partial \varepsilon_{\bm k}} \right)
+
D\left[
- \frac{e}{\hbar } \left( \frac{1}{\hbar} \frac{\partial \varepsilon_{\bm k}}{\partial \bm k} \times \bm B \right)
\right]
 \cdot 
\left(   \frac{\partial g_{\bm k}}{\partial \bm k}\right) 
= - \frac{g_{\bm k}}{\tau }.
\end{equation}

We use the ansatz 
\begin{equation}
g_{\bm k}= 
-e\tau D \left[\frac{e}{\hbar } (\bm B \cdot \bm E) (\bm \Omega \cdot \bm v) + \bm \Upsilon \cdot \bm v \right]  \left( -\frac{\partial f_0}{\partial \varepsilon_{\bm k}} \right)
\end{equation}
where $\bm \Upsilon$ is an unknown field independent of $\bm k$. The equation then reads
\begin{equation}
\label{}
(-e\bm E\cdot \bm v)
\left( \frac{\partial f_0}{\partial \varepsilon_{\bm k}} \right)
+
e\tau 
\left[
- \frac{e}{\hbar } \left( \bm v \times \bm B \right)
\right]
\cdot 
\left[ \frac{\partial  }{\partial \bm k}  \left(\frac{e}{\hbar } (\bm B \cdot \bm E)(\bm \Omega \cdot D\bm v  )
+
\bm \Upsilon \cdot D\bm v \right) \left( \frac{\partial f_0}{\partial \varepsilon_{\bm k}} \right)
\right]
= -e\bm v \cdot \bm \Upsilon \left( \frac{\partial f_0}{\partial \varepsilon_{\bm k}} \right).
\end{equation}

We change the cross product in the second term, in order to single out the velocity. Then we get
\begin{equation}
\label{}
(-e\bm E\cdot \bm v)
\left( \frac{\partial f_0}{\partial \varepsilon_{\bm k}} \right)
+e\tau 
\left[
- \frac{e}{\hbar } \bm B \right] \times 
\left[  \left(\frac{e}{\hbar } (\bm B \cdot \bm E) \nabla_{\bm k} (\bm \Omega \cdot D\bm v  )
+
\Upsilon_i \nabla_{\bm k} D v_i \right) \left( \frac{\partial f_0}{\partial \varepsilon_{\bm k}} \right)
\right]\cdot \bm v
= -e\bm v \cdot \bm \Upsilon \left( \frac{\partial f_0}{\partial \varepsilon_{\bm k}} \right)
\end{equation}

The variation of the equilibrium distribution and the electric charge are a common factor on both sides, then they can be eliminated. We get 
\begin{equation}
\label{}
\bm E +
\tau 
\left[
\frac{e}{\hbar } \bm B \right] \times 
\left[  \left(\frac{e}{\hbar } (\bm B \cdot \bm E) \nabla_{\bm k} (\bm \Omega \cdot D\bm v  )
+
\Upsilon_i \nabla_{\bm k} D v_i \right) 
\right] 
= \bm \Upsilon 
\end{equation}

We can define the vector $\bm C=\nabla_{\bm k} (\bm \Omega \cdot D \bm v  )$. The equation then reads
\begin{equation}
\label{}
\bm E 
+
\tau  \frac{e^2}{\hbar^2 }  (\bm B \cdot \bm E) \bm B\times \bm C
+
\tau D\frac{e}{\hbar} 
\bm B \times 
\left[
\Upsilon_i \nabla_{\bm k}D v_i 
\right] 
= \bm \Upsilon 
\end{equation}

The equation can be written as
\begin{equation}
\label{}
\bm E 
+
\tau  \frac{e^2}{\hbar^2 }  (\bm B \cdot \bm E) \bm B\times \bm C
+
\tau  \frac{e}{\hbar}  M_{L}  \bm \Upsilon
= \bm \Upsilon 
\end{equation}

where the matrix $ M_{L}$ reads (for notation we omit the factor $D$ in front of the velocity)
\begin{equation}
M_{L}=
\begin{pmatrix}
B_y\frac{\partial v_x}{\partial k_z}-B_z\frac{\partial v_x}{\partial k_y} 
& B_y\frac{\partial v_y}{\partial k_z}-B_z\frac{\partial v_y}{\partial k_y} 
& B_y\frac{\partial v_z}{\partial k_z}-B_z\frac{\partial v_z}{\partial k_y} \\
B_z\frac{\partial v_x}{\partial k_x}-B_x\frac{\partial v_x}{\partial k_z} 
& B_z\frac{\partial v_y}{\partial k_x}-B_x\frac{\partial v_y}{\partial k_z} 
& B_z\frac{\partial v_z}{\partial k_x}-B_x\frac{\partial v_z}{\partial k_z} \\
B_x\frac{\partial v_x}{\partial k_y}-B_y\frac{\partial v_x}{\partial k_x} 
& B_x\frac{\partial v_y}{\partial k_y}-B_y\frac{\partial v_y}{\partial k_x} 
& B_x\frac{\partial v_z}{\partial k_y}-B_y\frac{\partial v_z}{\partial k_x} \\
\end{pmatrix}.
\end{equation}

Then it follows 
\begin{equation}
\label{}
\bm \Upsilon = \left(I - \tau  \frac{e}{\hbar} M_{L} \right)^{-1}
\left[
\bm E 
+
\tau  \frac{e^2}{\hbar^2 }  (\bm B \cdot \bm E) \bm B\times \bm C
\right] 
\end{equation}

We can take the electric field as a common factor by writing the equation as
\begin{equation}
\label{}
\bm \Upsilon = \left( I - \tau  \frac{e}{\hbar} M_{L} \right)^{-1}
\left[
I_{3\times 3}
+
\tau  \frac{e^2}{\hbar^2 }M_{\nabla}
\right] \bm E 
\end{equation}

with the matrix 
\begin{equation}
M_{\nabla}=  (\bm B\times \bm C ) \otimes \bm B
=
\begin{pmatrix}
(\bm B\times \bm C )_x B_x & (\bm B\times \bm C )_x B_y & (\bm B\times \bm C )_x B_z \\ 
(\bm B\times \bm C )_y B_x & (\bm B\times \bm C )_y B_y & (\bm B\times \bm C )_y B_z \\ 
(\bm B\times \bm C )_z B_x & (\bm B\times \bm C )_z B_y & (\bm B\times \bm C )_z B_z 
\end{pmatrix}.
\end{equation}

and finally we can find the distribution 
\begin{equation}
g_{\bm k}= -e\tau D \left[-\frac{e}{\hbar } (\bm B \cdot \bm E) ( \bm \Omega \cdot \bm v ) - \bm v \cdot \bm \Upsilon  \right] \left( \frac{\partial f_0}{\partial \varepsilon_{\bm k}} \right). 
\end{equation}

In order to put every element in terms of a field proportional to the electric field, we can define a new matrix 
\begin{equation}
M_{\Omega}= \bm \Omega \otimes \bm B
=
\begin{pmatrix}
\Omega_x B_x & \Omega_x B_y & \Omega_x B_z\\
\Omega_y B_x & \Omega_y B_y & \Omega_y B_z \\
\Omega_z B_x & \Omega_z B_y & \Omega_z B_z
\end{pmatrix}.
\end{equation}

We define a new field $\bm \Phi=\frac{e}{\hbar } M_{\Omega}\bm E$ and can write the distribution as
\begin{equation}
g_{\bm k}= -e\tau D \left[-\bm v  \cdot \bm \Phi - \bm v \cdot \bm \Upsilon \right] \left( \frac{\partial f_0}{\partial \varepsilon_{\bm k}} \right). 
\end{equation}

The first term takes into account a $\bm k$-dependent driving field $\bm \Phi$. 
The second term contains $\bm \Upsilon$ 
which takes into account the Lorentz force and corrections due to the Berry curvature and its variations. This is explicit in the 
vector $\bm C=\nabla_{\bm k} (\bm \Omega \cdot D\bm v)$. 

In order to find the current we take the trace $j_i = -e \sum_{\bm k} \dot{\bm r}_{i} g_{\bm k}$. In changing the sum to an integral, we take into account that 
the density of states is corrected by the presence of the external magnetic field. Then, the rule is 
$\sum_{\bm k} \rightarrow \int \frac{ d^3\bm k }{(2\pi)^3} \left( 1+\frac{e}{\hbar} \bm B \cdot \bm \Omega \right) $. With this and using Eq.\eqref{Eq:velocity} the current follows as
\begin{align}
j_{i}
&=
e^2\tau 
\int \frac{ d^3\bm k }{(2\pi)^3} 
\left( 1+\frac{e}{\hbar} \bm B \cdot \bm \Omega \right)^{-1}
\left( \bm v  \cdot \bm \Phi + \bm v \cdot \bm \Upsilon \right)
\left(
\frac{1}{\hbar} \frac{\partial \varepsilon_{\bm k}}{\partial \bm k} + \frac{e}{\hbar }\bm E \times \bm \Omega
+
\frac{e}{\hbar } \left( \frac{1}{\hbar} \frac{\partial \varepsilon_{\bm k}}{\partial \bm k} \cdot \bm \Omega \right) \bm B
\right)
\left(-\frac{\partial f_0}{\partial \varepsilon_{0\bm k}} \right).
\end{align}

So far the equation is general. In order to calculate the longitudinal current contribution, we set the electric field in the direction of the velocity. Also, we set the magnetic field in the same direction as the electric field since we are interested in longitudinal magnetoresistance. In this case, the term $\propto \frac{e}{\hbar }\bm E \times \bm \Omega$ will not contribute since it is transversal and gives rise to the anomalous Hall effect. Also, $\bm \Gamma = \bm E$, since the other terms are transversal and give rise to the Lorentz force. The current can be written as
$j_{i}= \sigma_{ii}E_i$, where the conductivity reads
\begin{align}
\sigma_{ii}
&=
e^2\tau 
\int \frac{ d^3\bm k }{(2\pi)^3} 
\left( 1+\frac{e}{\hbar} \bm B \cdot \bm \Omega \right)^{-1}
\left[
v_i + \frac{e}{\hbar } \left(\bm v \cdot \bm \Omega \right) B_i
\right]^2
\left(-\frac{\partial f_0}{\partial \varepsilon_{0\bm k}} \right).
\end{align}

Expanding the correction of the density of states up to the second order in the magnetic field we get 
\begin{align}
\left( 1+\frac{e}{\hbar} \bm B \cdot \bm \Omega \right)^{-1}
\approx
1
-\frac{e}{\hbar} \bm B \cdot \bm \Omega 
+\left( \frac{e}{\hbar} \bm B \cdot \bm \Omega \right)^2.
\end{align}

The zeroth order term in magnetic field is the Drude conductivity
\begin{align}
\sigma_{ii,0}
&=e^2\tau  
\int \frac{ d^3\bm k }{(2\pi)^3} 
v^2_{i,0} \delta(\varepsilon_F - \varepsilon_{0\bm k} ) 
\end{align}

The first order in magnetic field reads
\begin{align}
\sigma_{ii,1}
&=
e^2\tau 
\int \frac{ d^3\bm k }{(2\pi)^3} 
\left[
2v_{i} \frac{e}{\hbar } \left(\bm v \cdot \bm \Omega \right) B_i
-\frac{e}{\hbar} \bm B \cdot \bm \Omega  v^2_i 
\right]
\left(-\frac{\partial f_0}{\partial \varepsilon_{0\bm k}} \right).
\end{align}

This term will vanish in our systems since we need to add the bands and the Berry curvature changes sign. 
The second order term in magnetic field reads
\begin{align}
\sigma_{ii,2}
&=
e^2\tau 
\left(  \frac{e}{\hbar } \right)^2
\int \frac{ d^3\bm k }{(2\pi)^3} 
\left[
v^2_{i} \left(  \bm B \cdot \bm \Omega \right)^2
-2v_{i}  \bm B \cdot \bm \Omega \left(\bm v \cdot \bm \Omega \right) B_i
+\left(\bm v \cdot \bm \Omega \right)^2 B^2_i
\right]
\left(-\frac{\partial f_0}{\partial \varepsilon_{0\bm k}} \right).
\end{align}


\subsection{Longitudinal magnetoresistance for the generalized Zeeman coupling of an ${\mathcal O}_h$ crystal }

%
In particular we get
\begin{align}
\sigma_{xx}(\bm B=0)
&=e^2\tau  
\left( \frac{2m}{\hbar^2} \right)
\left( \frac{\hbar}{m} \right)^2
\frac{1}{(2\pi)^3} \int d\phi \int \sin(\theta)d\theta \int dk k^4 \sin^2(\theta)\cos^2(\phi) \frac{\delta( k_F - k )}{2k_F} \\
&=e^2\tau  
\left( \frac{2m}{\hbar^2} \right)
\left( \frac{\hbar}{m} \right)^2
\frac{4\pi}{3}
\frac{1}{(2\pi)^3} \left( \frac{k^3_F}{2} \right)
\end{align}
After including spin degeneracy we find the usual expression $\sigma_{xx}(0)=n_{e}e^2\tau /m$ with $n_e =k^3_F/3\pi^2$.

For a magnetic field in the $y$-direction we get the second order correction
\begin{align}
\label{}
\sigma_{yy,2}(B_y)
&=
B^2_y e^2\tau  
\left( \frac{e}{\hbar } \right)^2 
\int \frac{ d^3\bm k }{(2\pi)^3} 
(v_x\Omega_x+ v_z\Omega_z)^2 
\delta(\varepsilon_F - \varepsilon_{0\bm k} )
\end{align}

We can calculate the Berry curvature as
\begin{equation}
\Omega^{+}_x
=\frac{1}{2}a^2b\frac{k_y k_x}{ [b^2+k^2_ya^2(k^2_x+k^2_z) ]^{3/2} }
=\frac{a^2}{2b^2}\frac{k_y k_x}{ \left[ 1+ k^2_y \frac{a^2}{b^2} (k^2_x+k^2_z) \right]^{3/2} }
\end{equation}
\begin{equation}
\Omega^{+}_z=
\frac{1}{2}a^2b\frac{k_y k_z}{ [b^2+k^2_ya^2(k^2_x+k^2_z) ]^{3/2} }
=\frac{a^2}{2b^2}\frac{k_y k_z}{ \left[ 1+ k^2_y \frac{a^2}{b^2} (k^2_x+k^2_z) \right]^{3/2} }
\end{equation}

To zeroth order in magnetic field the velocity reads $v_{x,0}= \hbar k_{x}/m $ and $v_{z,0}=\hbar k_{z}/m$. In the single relaxation time approximation we can write the 
conductivity as
\begin{align}
\label{}
\sigma_{yy,2}(B_y)
&=
2B^2_y e^2\tau  
\left( \frac{e}{\hbar } \right)^2 
\int \frac{ d^3\bm k }{(2\pi)^3} 
(v_{x,0}\Omega_x + v_{z,0}\Omega_z)^2 
\delta(\varepsilon_F - \varepsilon_{0\bm k} ).
\end{align}

The factor $2$ in front of the conductivity is because we need to sum over both bands, and both give the same result. 
The factor in parenthesis reads
\begin{align}
\label{}
(v_x\Omega_x + v_z\Omega_z)^2 =
\left(\frac{a^2}{2b^2} \right)^2  \left( \frac{\hbar}{m} \right)^2
\frac{1}{ \left[1+k^2_y\frac{a^2}{b^2}(k^2_x+k^2_z) \right]^{3}}k^2_y (  k^2_{x} + k^2_{z} )^2.
\end{align}

Then the integral reads
\begin{align}
\label{}
\Delta \sigma_{yy,2}(B_y)
&=
2B^2_y e^2\tau  
\left( \frac{e}{\hbar } \right)^2
\left(\frac{a^2}{2b^2} \right)^2 
\left( \frac{\hbar}{m} \right)^2
\frac{1}{(2\pi)^3}
\int d^3\bm k
\frac{ k^2_y ( k^2_{x} +  k^2_{z} )^2\delta(\varepsilon_F - \varepsilon_{0\bm k} )  }{ 
\left[ 1+k^2_y \frac{a^2}{b^2} (k^2_x + k^2_z ) \right]^{3}}
\end{align}

In spherical coordinates the integral reads
\begin{align}
\label{}
\int^{2\pi}_{0} d\phi
\int^{\pi}_{0} \sin(\theta)d\theta 
\int^{\infty}_0 k^2 dk 
\frac{ k^6 \sin^2(\theta)\sin^2(\phi)  ( \sin^2(\theta)\cos^2(\phi)  + \cos^2(\theta) )^2 }
{ \left[1+  k^4 \frac{a^2}{b^2}\sin^2(\theta)\sin^2(\phi) \left(\sin^2(\theta)\cos^2(\phi) + \cos^2(\theta) \right) \right]^{3}} 
\frac{\delta( k - k_F)}{2k_F} 
\end{align}

The integral over momentum is easily calculated. Defining the integral
\begin{align}
\label{}
I(k_F)= 
\int^{2\pi}_{0} d\phi
\int^{\pi}_{0} \sin(\theta)d\theta 
\frac{\sin^2(\theta)\sin^2(\phi)  ( \sin^2(\theta)\cos^2(\phi)  + \cos^2(\theta) )^2 }
{ \left[1 +  k^4_F\frac{a^2}{b^2} \sin^2(\theta)\sin^2(\phi) \left( \sin^2(\theta)\cos^2(\phi) + \cos^2(\theta) \right) \right]^{3}} 
\end{align}

the conductivity can be written as
\begin{align}
\label{}
\Delta \sigma_{yy,2}(B_y)=
2B^2_y e^2\tau  
\left( \frac{e}{\hbar } \right)^2
\left(\frac{a^2}{2b^2} \right)^2 
\left( \frac{\hbar}{m} \right)^2
\frac{1}{(2\pi)^3} 
\left( \frac{2m}{\hbar^2} \right)
\frac{k^7_F}{2}
I(k_F).
\end{align}

Using the definition of the Drude conductivity we find
\begin{align}
\label{}
\frac{ \Delta \sigma_{yy,2}(B_y)}{\sigma_{yy}(0)} 
=
\left( \frac{eB_y}{\hbar } \right)^2
\left(\frac{a^2}{2b^2}\right)^2 
\frac{3}{2\pi} k^4_FI(k_F).
\end{align}

The first factor defines the magnetic length or momentum as $1/l_{B} = k_{B}=\sqrt{eB/\hbar}$. We define a characteristic length  
related to the material properties as $l_{0}=\sqrt{a/b}$ or conversely a characteristic momentum $k_0=1/l_0$. Then one writes 
\begin{align}
\label{}
\frac{ \Delta \sigma_{yy,2}(B_y)}{\sigma_{yy}(0)} 
&= \left( \frac{l_0}{l_B} \right)^4 \frac{3}{8\pi} \left( \frac{k_F}{k_0}\right)^4 I( k_F/k_0) 
\end{align}

The integral reads
\begin{align}
\label{}
I(k_F/k_0)= 
\int^{2\pi}_{0} d\phi
\int^{\pi}_{0} \sin(\theta)d\theta 
\frac{\sin^2(\theta)\sin^2(\phi)  ( \sin^2(\theta)\cos^2(\phi)  + \cos^2(\theta) )^2 }
{ \left[1 +  \left( k_F l_{0} \right)^4 \sin^2(\theta)\sin^2(\phi) \left( \sin^2(\theta)\cos^2(\phi) + \cos^2(\theta) \right) \right]^{3}}.
\end{align}


\section{Generalized Zeeman coupling  in centrosymmetric crystals}

To establish the appearance of a Berry curvature and the form of the generalized Zeeman coupling, we note that the latter has to contain terms that are even in the crystalline momentum, due to inversion symmetry. 
In the absence of additional symmetries the most general 
coupling up to the second order in momentum will be of the form ${\mathcal H}_{z}=\bm{d} \cdot \boldsymbol \sigma$ with the ${\bm d}$ vector that can be expressed as
\begin{equation}
\bm d = (a_0 + a_ik^2_i + a_{ij}k_{i}k_{j}  )\hat{\bm{\imath} } + (b_0 + b_ik^2_i + b_{ij}k_{i}k_{j} )\hat{\bm{\imath} } + (c_0 + c_ik^2_i + c_{ij}k_{i}k_{j})\hat{\bm{z} }
\end{equation}
where the coefficients $a_{ij},b_{ij}$ and $c_{ij}$ we have $i\neq j$. 
The presence of residual (magnetic) point group symmetries pose strong constraints on the coefficients of the binomials. To establish the presence of the Zeeman-activated Berry curvature in all centrosymmetric crystals, we therefore derive the generalized Zeeman coupling 
in all magnetic point groups compatible with ferromagnetism -- in our case this is due to the presence of the homogeneous external magnetic field. 
The table below explicitly show the form of the generalized Zeeman coupling. The direction of the magnetic field is specified by the non-vanishing term independent of momentum, {\it i.e.} the conventional Zeeman coupling.

\begin{table}[ht]
\centering
\caption{Effective Zeeman coupling allowed in centrosymmetric magnetic groups compatible with ferromagnetism. }
\begin{tabular}[t]{||c||c||c||c||}
\hline
Magnetic group & $d_x$ & $d_y$ & $d_z$  \\
\hline \hline
 $ 6/mm'm' $ & $a_{xy}k_xk_y $ & $b_0 + b_x(k^2_x + k^2_z) + b_y k^2_y$ &  $a_{xy}k_xk_z$ \\  
\hline
$ 6/m $  & $a_{xy}k_xk_y $ & $b_0 + b_x(k^2_x + k^2_z) + b_y k^2_y$ &  $ a_{xy}k_xk_z$ \\  
\hline
$ -3m'$ & $a_{xy}k_xk_y $ & $b_0 + b_x(k^2_x + k^2_z) + b_y k^2_y$ &  $ a_{xy}k_xk_z$ \\  
\hline
$-3$ & 
\makecell[l]{
$ a_{xz}k_xk_z - a_{yz}k_yk_z + a_{x}(k^2_x -k^2_y) $\\
$ - 2a_{xy}k_xk_y $
} & 
\makecell[l]{
$ a_{xz}k_yk_z +a_{yz}k_xk_z -2a_{x} k_xk_y $\\
$ - a_{xy}(k^2_x - k^2_y) $
} & 
$ c_0 + c_x(k^2_x + k^2_y)+ c_zk^2_z $\\
\hline
 $ 4/mm'm' $ & $a k_xk_y$ & $b_0 + b_{x}(k^2_x + k^2_z) + b_y k^2_y$ & $a k_zk_y$ \\ 
 \hline
$ 4/m$  & $a_{xy}k_xk_y + a_{yz}k_yk_z $ & $b_0 + b_{x}(k^2_x + k^2_z) + b_y k^2_y$ &  $-a_{yz}k_xk_y + a_{xy}k_yk_z $  \\ 
\hline
$ m'm'm $  & $a_{xy}k_xk_y$ & $b_0 + b_{i}k^2_i$ & $c_{xz}k_y k_z$ \\ 
\hline
$ 2'/m'$  & $a_0 + a_{i}k^2_i + a_{zx}k_{z}k_{x} $ & $ b_{xy}k_{x}k_{y} + b_{zy}k_{z}k_{y} $ & $ c_0 +  c_{i}k^2_i + c_{zx}k_{z}k_{x} $  \\ 
\hline
$ 2/m$ & $ a_{xy}k_{x}k_{y} + a_{zy}k_{z}k_{y} $ & $b_0 + b_ik^2_i + b_{xz}k_{x}k_{z}$ & $ c_{xy}k_{x}k_{y} + c_{zy}k_{z}k_{y}$   \\ 
\hline
$ -1 $  & $a_0 + a_ik^2_i + a_{ij}k_{i}k_{j} $ & $b_0 + b_ik^2_i + b_{ij}k_{i}k_{j} $ & $c_0 + c_ik^2_i + c_{ij}k_{i}k_{j} $  \\ 
\hline \hline
\end{tabular}
\end{table}

\end{document}